\documentclass[prl,twocolumn,floatfix,showpacs]{revtex4-1}

\usepackage{epsf,amsmath,amsthm,amssymb,amsfonts,amsbsy}
\usepackage{tikz}
\usetikzlibrary{arrows}

\usepackage{setspace}

\usepackage{dcolumn}   
\usepackage{bm}        
\usepackage{subfigure}

\usetikzlibrary{arrows}

\DeclareGraphicsExtensions{.pdf,.jpg}

\newcommand{\norm}[1]{\lVert #1 \rVert}

\newcommand{\bu}{{\bf u}}

\newcommand{\grad}{\nabla}

\renewcommand{\phi}{\varphi}

\newcommand{\beq}{\begin{eqnarray}}
\newcommand{\eeq}{\end{eqnarray}}

\newcommand{\pt}{\partial}

\newcommand{\tilu}{\widetilde{u}}

\theoremstyle{remark}

\DeclareGraphicsExtensions{.pdf}

\begin{document}

\title{Spatial Pattern Dynamics due to the Fitness Gradient Flux in Evolutionary Games}

\author{Russ deForest
and Andrew Belmonte}\altaffiliation{Also at Dept of Materials Science \& Engineering, Pennsylvania State University}\email{andrew.belmonte@gmail.com}

\affiliation{The W.~G.~Pritchard Laboratories, Department of Mathematics,
Pennsylvania State University, University Park, PA 16802, USA}


\date{December 12, 2012}

\begin{abstract}
We introduce a non-diffusive spatial coupling term into the replicator
equation of evolutionary game theory. The spatial flux is based on motion due to local gradients in the relative fitness of each strategy, providing a game-dependent alternative to diffusive coupling. 
We study numerically the development of patterns in 1D for two-strategy
games including the coordination game and the prisoner's dilemma, and in 2D
for the rock-paper-scissors game. 
In 1D we observe modified travelling wave solutions in the presence of diffusion, and asymptotic attracting states under a frozen strategy assumption without diffusion. In 2D we observe spiral formation and breakup in the frozen strategy rock-paper-scissors game without diffusion. A change of variables appropriate to replicator dynamics is shown to correctly capture the 1D asymptotic steady state via a nonlinear diffusion equation.


\end{abstract}




\maketitle



Evolutionary games provide a convenient and promising method for
modeling the dynamics of selection and competition in biological and social
processes \cite{msmithBOOK,nowakBOOK}. While game theory at its origin focused on individual strategic decisions in a game played once \cite{vonBOOK}, time was later introduced in terms of members of a population who play repeatedly, and change their strategy according to some rule \cite{msmithBOOK}, or equivalently in terms of changes in the frequency of inherited traits or the genes that represent them \cite{fisher37}. This dynamic approach has already shed light on how group interactions can support traits that are not obviously advantageous \cite{smith73,nowak06}.

One way to mathematically model the evolution of frequencies of a given strategy within a population is via the replicator equation \cite{taylor78,schuster83}.
For strategies $i = 1, ..., m,$ the replicator dynamics for the $i$th strategy are defined by 
 \begin{equation}
     \label{eq:1}
     \frac{d u_i}{dt} = u_i\left[f_i(\bu) -\overline{f(\bu)}\right],
 \end{equation}
where $\bu(t) = (u_1, ..., u_m)$ denotes the frequencies (number of players) of each strategy in the population with $\sum u_i = 1$. 
The fitness of strategy $i$ is written $f_i(\bu)$, and the average fitness of the entire population is $\overline{f(\bu)}$. In evolutionary games, the
fitness of a strategy depends not only on the payoff for that
strategy, but also on the frequencies of every other strategy in the population. If $A$ is the $m \times m$ payoff matrix for a symmetric game, then a natural choice for the fitness function is $f_i = (A\bu)_i$, the expected payoff for an individual playing strategy $i$, and $\overline{f} = \bu^TA\bu$ is the scalar average payoff for all strategies.




Diffusive spatial coupling or random mobility in evolutionary games allows for pattern formation and gives rise to interesting and rich dynamics; it can also support outcomes not attained without the inclusion of spatial effects \cite{vickers89,ferriere00,sicardi09}. For example, an evolutionarily stable strategy (ESS) is one which cannot be invaded by another strategy at an initially small frequency or population level \cite{msmithBOOK}. However, it has been shown in a spatial setting that an ESS can invade and replace another ESS via a diffusion-induced traveling wavefront \cite{hutson95}. In public goods games, unequal diffusion of strategies can promote cooperative behavior by allowing the productive cooperators to aggregate and coexist among the more rapidly diffusing freeloaders,
similar to a Turing instability \cite{wakano09,wakano11}.
Other approaches to introducing spatial dimensions into evolutionary games include fixed individuals on a grid playing against nearest neighbors \cite{nowak92,killingback96}, or against randomly-chosen neighbors while allowing for position swaps \cite{reichenbach07}.
The inclusion of diffusive coupling in the replicator ODEs corresponds to an assumption that all individuals in the population wander randomly while changing strategies. This means that
the local flux of those playing the $i$th strategy is given by Fick's Law: 
$J_i = -D_i\grad{u_i}$, with $D_i$ the diffusion constant. What results is  essentially a reaction-diffusion system, where the ``reaction" comes from the competitive interaction between players \cite{vickers89,ferriere00}.




Once space is included, the frequency interpretation of ${\bf u}$ no
longer holds, because $\sum u_i$ varies from point to point; we can still treat ${\bf u}(x,t)$ as the vector of local population densities, and $\norm{{\bf u}}_1 \equiv \sum u_i$ the total density (number of players) at $(x,t)$. 
Following the approach taken in diffusive evolutionary games by Vickers \cite{vickers89}, we define the fitness $f_i({\bf u}) = (A{\bf u})_i /\norm{{\bf u}}_1$, the expected payoff for strategy $i$, and the average payoff
$\overline{f(\bu)} = \bu^T A\bu / \norm{{\bf u}}_1^2$.


\smallskip

What if the players do not move randomly?
A more general transport law for spatial evolutionary games is given by the following expression for the flux 
\begin{equation}
  \label{E_newflux}
  J_i = -D_i\grad u_i + \beta_i u_i\grad\left[(A{\bf u})_i /\norm{{\bf u}}_1\right].
\end{equation}
Here $\beta_i$ is the proportionality constant for the fitness gradient $\grad f_i$, which allows for the spatial movement of players in the direction of increasing payoff. Including a flux of players moving in a profitable direction, as an alternative to changing strategies, is a sort of spatial version of the replicator equation time dynamics. 
The same approach has been taken in modeling dispersal and spatial distribution in ecology \cite{shigesada79, cosner05}; a related idea was introduced as ``success-driven migration" in a prisoner's dilemma model on a grid \cite{helbing09}, which has recently been extended to model reputation-based migration \cite{cong12}. 
Here we use the flux in Eq.~\ref{E_newflux} to obtain the partial differential equation (PDE) system:
$$
  \frac{\partial u_i}{\partial t} =
  \alpha_i u_i\left[(A{\bf u})_i /\norm{{\bf u}}_1 
  -\bu^T A\bu / \norm{u}_1^2\right] \qquad \qquad
$$
\vspace{-7mm}
\begin{equation}
\qquad \qquad \qquad 
-\beta_i\nabla\cdot (u_i\nabla[ (A{\bf u})_i /\norm{{\bf u}}_1])
  +D_i \Delta u_i,
    \label{E_mainmodel}
\end{equation}
where $\alpha_i$ is a proportionality constant for the game.
The derivation of the fitness gradient flux begins with the assumption that the probability of a player moving is proportional to the local difference between the fitness at adjacent points in space \cite{Note}. In the continuum limit this contributes a term $\grad f_i$, which appears in the flux multiplied by the local player density $u_i$, proportional to the number of players with fitness $f_i$.






Here we present a study of the consequences of this non-diffusive spatial coupling in several $2\times 2$ and $3\times 3$ symmetric games. Throughout this paper we will use $\beta_1=\beta_2 = \beta$, in other words we will not study the possibility of different flux coefficients of response to the fitness gradient. Similarly we will take $\alpha_1=\alpha_2 = \alpha$ for the replicator term coefficients.  We will however allow for differences in diffusion, $D_1 \neq D_2$. For convenience, we will use the following notation for $2\times 2$ games: $u_1 = u, u_2 = v, D_1 = D_u$, etc. 
Our numerical results were obtained with a fully-implicit scheme using a parallelized Newton iteration, on spatial grids of 256 to 4048 points (1D) or 256 $\times$ 256 points (2D).
Semi-implicit schemes are frequently
employed in reaction diffusion systems, allowing the nonlinear reaction to be
handled explicitly \cite{Ruuth95}.  Since the spatial derivatives in our system are nonlinear, an implicit treatment of the nonlinearity cannot be avoided.

\begin{figure}[!b]
\centering
\includegraphics[width=8.9cm]{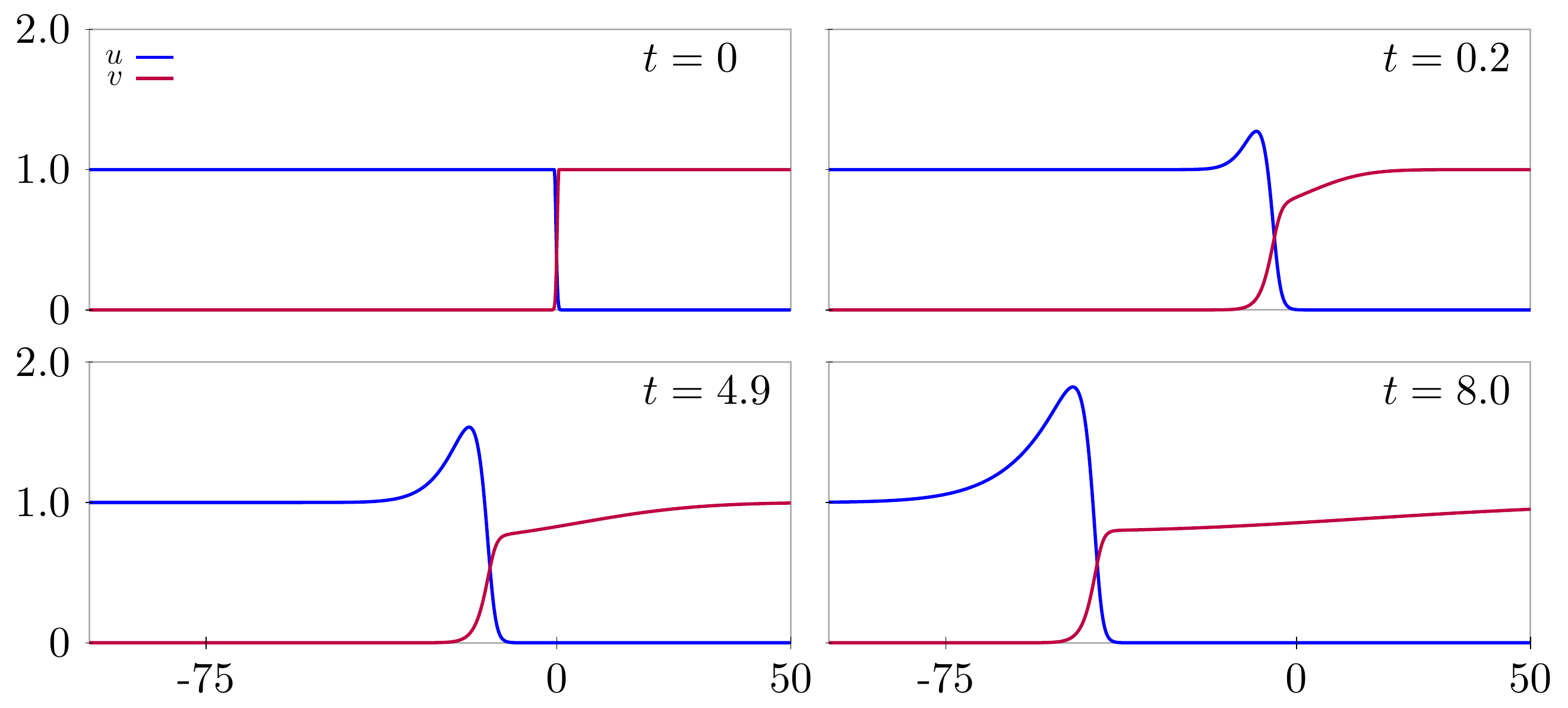} 
\caption{Spatial dynamics of a travelling wave in the adjacent region geometry for a coordination  game with payoff matrix given in text, $\alpha = 1$,  $D_u= 1$, $D_v=3$, and $\beta = 0.2$.  Times shown are as labelled.
}
\label{FIG_Vick1}
\end{figure}

To study the effects of the fitness gradient flux on previously known results, we compare with phenomena obtained by Hutson and Vickers in the setting of a pure coordination game in one dimension \cite{hutson95}; we find that the fitness gradient flux modifies their results strikingly. In a pure coordination game, strategies prefer their own type, as is evident from its payoff matrix $A =
\begin{bmatrix}
  a_{11} & 0\\ 0 & a_{22}
\end{bmatrix}$. 
If $a_{22} > a_{11}$, it would be more advantageous for the population to play the 2nd ($v$) strategy, nonetheless a population uniformly playing the $u$ strategy is known to be stable (ESS), and cannot be invaded by a small population playing the other strategy. In fact each pure strategy is evolutionarily stable \cite{Szabo2007}. However, Hutson and Vickers proved that diffusive coupling can produce travelling wavefronts, whereby a population $v$ invades $u$ when $a_{22} > a_{11}$ \cite{hutson95}. The initial conditions they considered were two adjacent regions of pure strategy dominance with no overlap (our initial conditions will be smooth functions with a small overlap, see e.g.~Fig.~\ref{FIG_Vick1}a). In this adjacent region geometry, diffusion initiates replicator dynamics as the sharp initial population distributions relax and overlap with time. Depending on the diffusion constants, a local increase or bump in $u$ ($D_v>D_u$) or $v$ ($D_u>D_v$) occurs at the region boundary in the travelling wave solution \cite{hutson95}.

The fitness gradient flux adds spatial gradient terms which may modify existence or other properties of these solutions. From a qualitative standpoint, the diffusive flux between the adjacent region can now be enhanced or opposed by the fitness gradient flux.  
Generally, the PDEs in Eq.~\ref{E_mainmodel} can be shown to be normally parabolic \cite{russ2}, and thus well-posed for $\beta$ values such that
\begin{equation}
0\leq \beta < \frac{4 D_u D_v}{D_v(a_{11}-a_{12})+D_u(a_{22}-a_{21})} \equiv \beta_c;
    \label{E_beta}
\end{equation}
in other words, for strong enough diffusion. The equation has smooth solutions in this range of $\beta$, but solutions blow up as $\beta$ approaches $\beta_c$ \cite{amann89,amann90}.

\begin{figure}[!t]
\centering
\includegraphics[width=8.5cm]{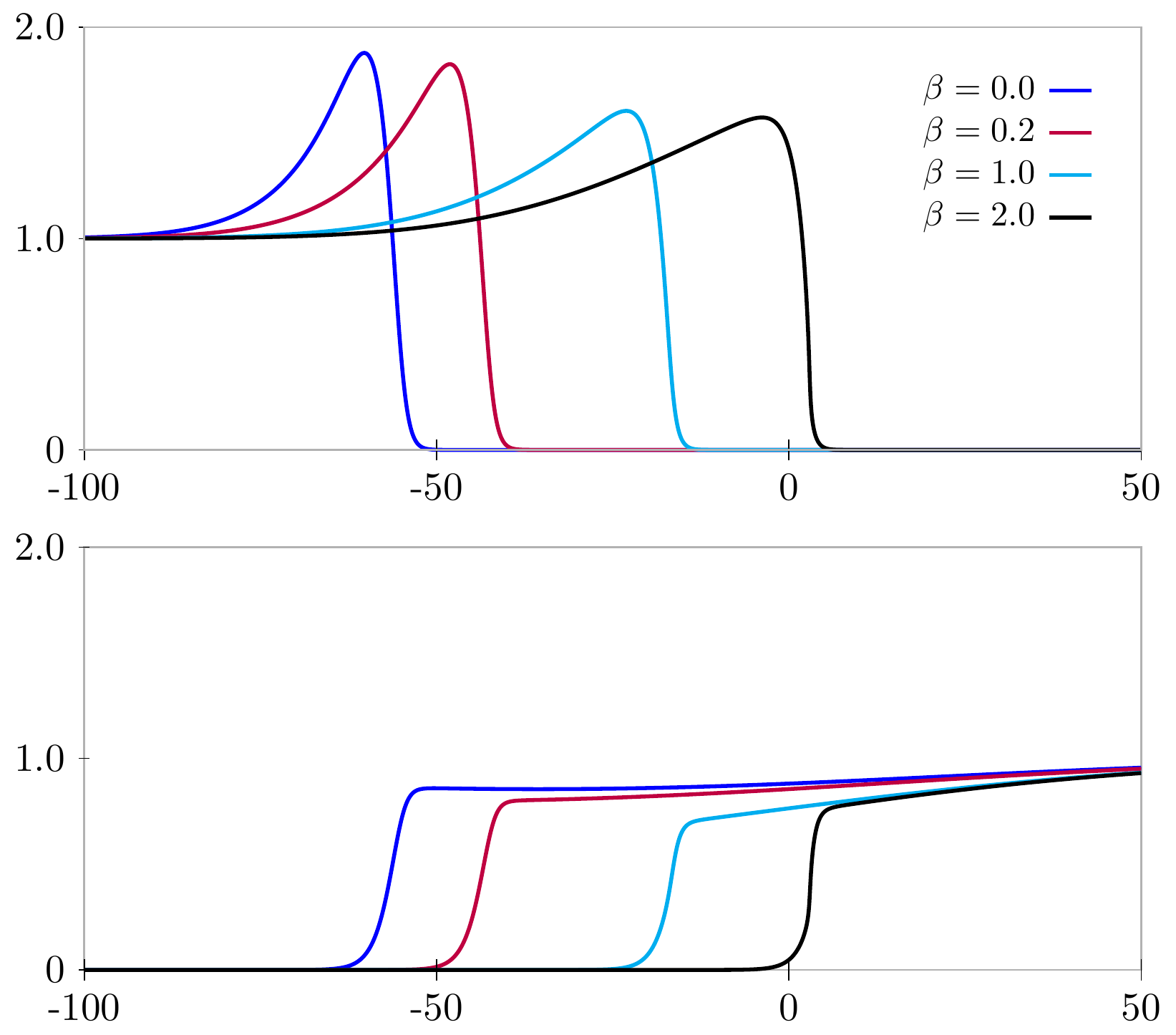} 
\put(-220,195){a)}
\put(-220,88){b)}
\caption{Effects of the fitness gradient flux on fully developed travelling wave solutions in the coordination game: a) $u(x,t)$, b) $v(x,t)$ at $t=390.2$, for initial conditions in the adjacent region geometry as in Fig.~\protect \ref{FIG_Vick1}. Here $D_u= 1$, $D_v=3$, and $\alpha = 1$; values of $\beta$ as labelled. Note that for both $\beta = 1$ and 2, the solutions have reached a steady state (speed $c=0$). 
}
\label{FIG_Vick2}
\end{figure}

We numerically simulate Eq.~\ref{E_mainmodel} for a coordination game with $D_u=1$, $D_v=3$, $\alpha = 1$,  and payoff matrix $A =
\begin{bmatrix}
  1 & 0\\ 0 & 2
\end{bmatrix}$,  
in which case $\beta_c = 2.4$. We study Hutson-Vickers solutions in an adjacent region geometry for $\beta<\beta_c$, and find that the travelling wave continues to exist for $\beta >0$, albeit with the observed speed of translation $c$ reduced. An example of the development of such a travelling wave solution is shown in Fig~\ref{FIG_Vick1}, for $\beta = 0.2$. Note that a bump is still observed in $u$, corresponding to the condition $D_v = 3 >D_u$ \cite{hutson95}. As $\beta$ is increased, the wave speed $c$ of these solutions appears to decreases linearly with $\beta$ (for $\beta=0.1-0.6$), and stops entirely for $\beta \ge 0.7$, see Fig~\ref{FIG_Vick2}. For those cases with $c=0$, the wavefront develops initially, but the invasion is halted and the solution achieves a steady state ($\beta = 1$ and 2 in Fig~\ref{FIG_Vick2}).

\begin{figure}[!b]
\centering




\def\Xshift{890}
\def\Yshift{430}
\tikzstyle{axes}=[thick,gray!60]
\tikzstyle{ticks}=[thick,gray!60]
\tikzstyle{uplot}=[thick,blue]
\tikzstyle{vplot}=[thick,purple]
\tikzstyle{uvplotscale}=[xscale=1,yscale=10]
\tikzstyle{plot1}=[xshift=0]
\tikzstyle{plot2}=[xshift=\Xshift]
\tikzstyle{plot3}=[xshift=0,yshift=-\Yshift]
\tikzstyle{plot4}=[xshift=\Xshift,yshift=-\Yshift]
\def\xmin{-20}
\def\xmax{10}
\def\ymin{0}
\def\ymax{1.4}
\def\yticks{0,0.5,1.0}
\def\xticks{-16,-10,0,6}
\def\textscale{.9}
\def\FIGDIR{./}


  \begin{tikzpicture}[scale=.13]


    \foreach \x/\xtxt/\time in {plot1/init/0, plot2/1/0.2,
      plot3/3/4.9, plot4/5/8.0}{



    \begin{scope}[\x,axes,uvplotscale]
      \draw (\xmin,\ymin) rectangle (\xmax,\ymax);
      \draw (0,1.25) node[right,color=black,scale=\textscale] {$t=\time$};     
    \end{scope}
   

    \begin{scope}[\x,uvplotscale]
      \draw[uplot] plot file {\FIGDIR u\xtxt.txt};
      \draw[vplot] plot file {\FIGDIR v\xtxt.txt};
    \end{scope}

}   

    \foreach \i in {plot1,plot3}{
        \begin{scope}[\i,uvplotscale]
    \foreach \j in \yticks {
        \draw[black] (-20.3,\j) node[left,scale=\textscale] {\j}--(-19.7,\j);
      }
  \end{scope}
}

  \foreach \i in {plot2,plot4}{
        \begin{scope}[\i,uvplotscale]
           \foreach \j in \yticks {
        \draw[black] (-20.3,\j) --(-19.7,\j);
      }
        \end{scope}
    
  }


 \foreach \i in {plot3,plot4} {
 \begin{scope}[\i,uvplotscale]
     \foreach \x in \xticks{
    \draw (\x,-.025) node[anchor=north,scale=\textscale] {\x}--(\x,.025);
 }
  \end{scope}
 }


 \begin{scope}[plot1,uvplotscale]
      \draw[uplot] (-14,.7) node[left,scale=1,color=black]
      {$u$}--(-12,.7);
      \draw[vplot] (-14,.45) node[left,scale=1,color=black] {$v$}--(-12,.45);
    \end{scope}
  \end{tikzpicture}


\caption{Spatial dynamics of a travelling wave in the adjacent region geometry for a prisoner's dilemma game with payoff matrix given in text, $\alpha = 1$, $D_u=D_v=0.1$, and $\beta = 1$.  Times shown are as labelled.
}
\label{FIG_pd1}
\end{figure}
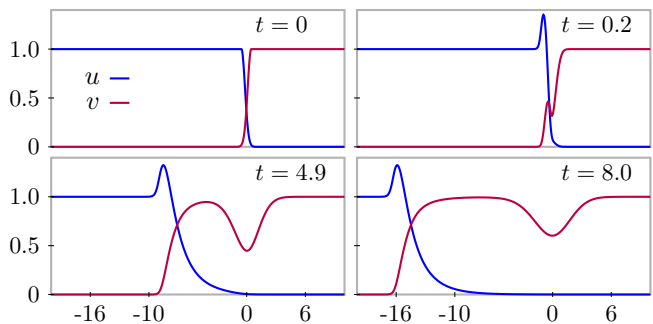	

Moreover, we observe 
similar travelling wave solutions numerically in the prisoner's dilemma game, which was not considered in \cite{hutson95}. An example is shown in Fig.~\ref{FIG_pd1}, for $A =
 \begin{bmatrix}
   1 & -1\\ 3 &0
 \end{bmatrix}
$
and $\beta = 1$. This is quite different from the purely diffusive coupling case ($\beta = 0$), for which no travelling waves are seen. Instead, the system progresses towards the only Nash equilibrium of the game (everyone defects, which is also an ESS) \cite{Szabo2007}, with dynamics essentially unmodified by the spatial coupling. The travelling wave which occurs with the inclusion of fitness gradient flux may be caused by the directed motion of cooperators, who can now flee in front of the takeover of defectors. We do not yet know if these observations for the prisoner's dilemma game will correspond analytically to the travelling wave solutions in the coordination game \cite{hutson95}.

We next consider the dynamics of 2-strategy games in one-dimension and study the fitness gradient flux in the no reaction, no diffusion case ($\alpha_i=0$ and $D_i=0$ in Eq.~\ref{E_mainmodel}). This corresponds to a situation in which players who cannot change strategy, but may nonetheless move in order to improve their benefit---their strategy is ``frozen''.  The dynamics still depend on the properties of the payoff matrix $A$ through the fitness gradient flux in Eq.~\ref{E_mainmodel}, which provides a game-dependent spatial coupling.

\begin{figure}[!t]
\centering
\includegraphics[width=8.55cm]{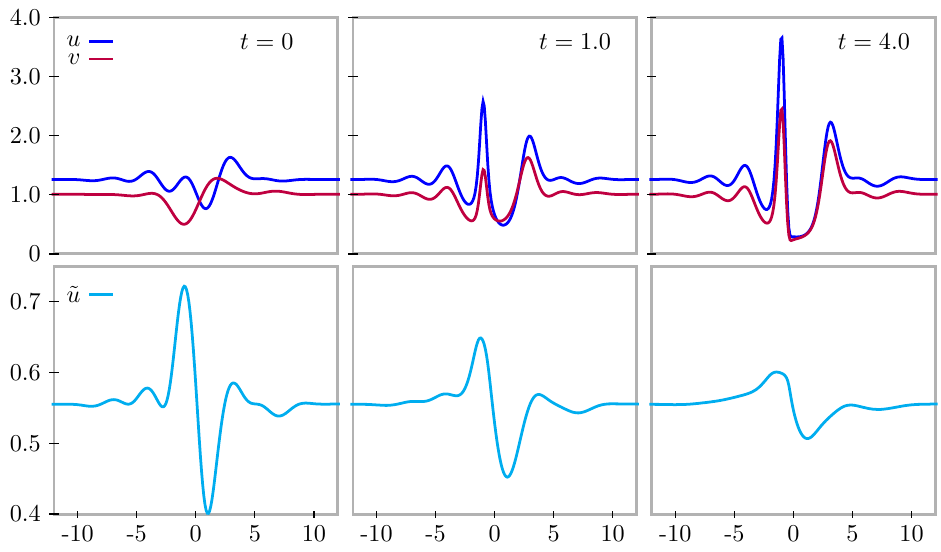} 
\hspace{1.5mm}
\vspace{-4mm}	 
\caption{Top row: 1D spatial dynamics of a random mode perturbation of $u$ and $v$ for a prisoner's dilemma game with no reaction or diffusion, showing the progress towards a structured steady state, times as labelled.
Bottom row: corresponding evolution of the proportionate variable $\tilu (x,t)$. Here $\kappa(A) = 1$ (see text).}
\label{FIG-uvseries}
\end{figure}

We find numerically that for certain classes of games, characterized by a simple condition on the matrix $A$ discussed below,  the frozen strategy
assumption leads to a spatially structured steady state.  An example
is shown in Fig.~\ref{FIG-uvseries} for random initial conditions in the prisoner's dilemma game,  $A =
\begin{bmatrix}
  1 & -1\\ 3 & 0
\end{bmatrix}
$. Note that this steady state does not represent the coexistence of opposing strategies in the usual sense (cooperation), because by assumption players are not allowed to change strategy.

%

To begin to understand this steady state, we let $u_1=u,\: u_2=v$, and define the proportionate variable
\begin{equation*}
\widetilde{u}(x,t) \equiv \frac{u(x,t)}{\norm{{\bf u}}_1} =  \frac{u(x,t)}{u(x,t) + v(x,t)},
\end{equation*}
with a similar definition for $\widetilde{v}(x,t)$. Because $\tilu + \widetilde{v}= 1$ at each point, only one of these variables is needed. However, we do not obtain a new PDE purely in terms of $\tilu$. Defining $p(x,t) = \nabla u /(u+v)$, we obtain
\begin{equation}
\label{E-nonlineardiff}
\pt_t \tilu = L(\tilu, p, A)\cdot\nabla \tilu +  \kappa(A) \, \tilde{D} \, 
\tilu (1 - \tilu) \Delta \tilu,
\end{equation}
where $L$ is a nonlinear
function, $\tilde{D} > 0$ is a constant, and for payoff matrix elements $a_{ij}$,
the constant 
$$
\kappa(A) \equiv a_{12} + a_{21} - (a_{11}+a_{22})
$$ 
controls the spatial dynamics \cite{russ2}. For $\kappa(A)<0,$ Eq.~\ref{E-nonlineardiff} is ill-posed as a backwards heat equation, which corresponds to cooperators aggregating faster than defectors can follow as the solutions blows up in finite time.
For cases with $\kappa(A)>0$, we find that solutions evolve to a steady state defined by $\nabla \tilu =0$, which defines a steady state solution to Eq.~\ref{E-nonlineardiff} - an example is shown in Fig.~\ref{FIG-uvseries}.

It is interesting to consider solutions to Eq.~\ref{E-nonlineardiff} in reference to the  classification of two-strategy matrix games proposed by Szab{\'o} \& F{\'a}th (Sect. 2.4 in \cite{Szabo2007}): 
\vspace{-1mm}

\begin{enumerate}
\item Anti-coordination class: games with $a_{11}<a_{21}$ and
  $a_{22}<a_{12}$. The coefficient $\kappa(A)>0$ for all games in this
  class and the system evolves to a steady state as each population
  locally adjusts to the relative benefit between its own type and the other.
\item Coordination class: games with $a_{11}>a_{21}$ and $a_{22}>a_{12}$, 
  so $\kappa(A)<0$.  Under the fitness gradient flux, populations will 
  seek to aggregate by strategy type. Well-posedness in this case can only occur in the presence
  of diffusion, with sufficiently high coefficients equivalent to the $\beta<\beta_c$ condition
  discussed above. 
\item Pure-dominance class: games with $(a_{11}-a_{21})(a_{22}-a_{12})<0$. The coefficient $\kappa(A)$ can be of either sign here, so well-posedness depends on the specific choice of $A$, sufficiently high diffusion, or other assumptions.
In the prisoner's dilemma, cooperators $u$ aggregate locally, while defectors $v$ follow them.
  If $A$ is chosen so that $\kappa(A)>0$, as in Fig.~\ref{FIG-uvseries}, then the system
  evolves toward a steady state - the local aggregation of $u$ into four peaks is 
  readily apparent in that case. For games with $\kappa(A)<0$, the PDE is ill-posed.
\end{enumerate}


\begin{figure}[!b]
\begin{center}
\includegraphics[width=8.7cm]{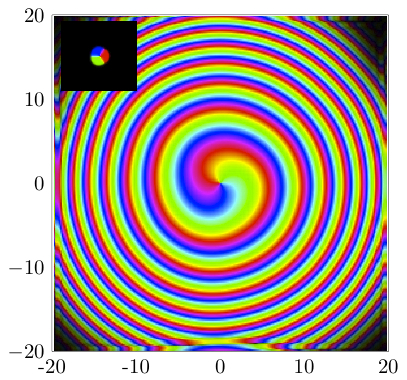} 
\vspace{-7mm}	
\caption{Spiral pattern in the frozen-strategy rock-paper-scissors game, showing its full extent at $t = 5.0$; strategy fields $u$, $v$, $w$ are shown in an additive RGB color map (see text). The inset shows a closeup of the initial conditions.}
\label{FIG_rps1}
\end{center}
\end{figure}

\begin{figure}[!t]
\begin{center}
\hspace{-1mm}	
\includegraphics[width=8.7cm]{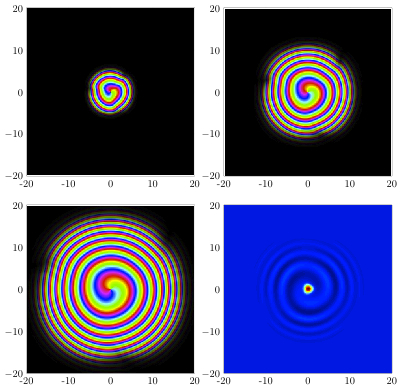} 
\caption{Development of a spiral pattern, for times  a) $t = 0.5$, b) 1.6, c) 3.0, with strategy fields shown in an additive RGB color map; d) corresponding heat map plot for one individual strategy $u_1(x,t)$ is shown, also at t = 3.0.}
\label{FIG_rps2}
\end{center}
\end{figure}

Three strategy games allow for the possibility of cyclic dominance of strategies \cite{hofBOOK}, where any given strategy can be beaten by another. We investigate the dynamics of the three strategy rock-paper-scissors game \cite{hofBOOK} under the frozen strategy assumption, defined by the payoff matrix
\begin{equation}   
A =
  \begin{bmatrix}
    0 & -1 & 1 \\ 1 & 0 & -1\\ -1 & 1 & 0
      \end{bmatrix}.
      \label{E-rps} 
\end{equation} 
We observe spiral waves in this system, as shown in Fig.~\ref{FIG_rps1} with periodic boundary conditions; spiral waves in the rock-paper-scissors games are reported elsewhere for other spatial couplings \cite{reichenbach07,jiang11}. Because it is difficult to represent all three strategy populations on a single plot, we use the following normalized version of the additive RGB color system. We first assign a color to each strategy; at each $(x,t)$, the $u, v, w$ values are normalized 
such that the minimum is 0, the maximum is 1, and the other is in the closed interval $[0,1]$. These numbers determine the RGB value for that point \cite{RGB}. The result is an image indicating which strategy is predominant, and the relative proportion of the other two (an example of a single strategy concentration field for a spiral wave is shown in  Fig.~\ref{FIG_rps2}d).

The initial conditions consist of a superposition of Gaussian bumps for each strategy population, of the form
\begin{equation}
  u_i({\bf x},0)=C_ie^{-K_i({\bf x}-{\bf x}_i)^2}+1,\quad i=1,2,3
  \label{E-bump}
\end{equation}
To produce a spiral pattern, the points ${\bf x}_i$ are placed symmetrically on a circle of radius $r$ about the origin \cite{russ2}. In Fig.~\ref{FIG_rps1}, $r=2$ with $K_i=5$ and $C_i=1$ for all $i$ (256 x 256, periodic boundary conditions, $\Delta t = 0.01$).  Spiral formation is determined by the proximity and magnitude of these bumps, 
via the parameters $r$ and $C_i$. For fixed $C_i$, there is a range $r_1< r< r_2$ where spiral formation is orderly and persists until boundary effects cause it to break up; this breakup is already evident at the edges of Fig.~\ref{FIG_rps1}. An example of spiral development is shown in  Fig.~\ref{FIG_rps2}a-c. The pattern rotates as the three Gaussian bumps chase each other, driven by the fitness gradient flux, and density waves are radiated forming the spiral arms. 
When the initial bumps are close (small $r$), the spiral forms right away, but breaks up before it reaches the boundaries. If the bumps are further apart
(large $r$), formation proceeds more slowly, and is incomplete before the pattern is complicated by boundary effects \cite{russ2}. 

To test the robustness of the spiral to asymmetric initial conditions, we varied the size $K_i$ of the Gaussian bump for each strategy, and the payoffs by replacing the $\pm 1$ in Eq.~\ref{E-rps} with values drawn from a uniform random distribution on $[0,1]$. 
While these changes introduced some asymmetries, they did not stop spiral formation from the three bump configuration.

Spiral waves are observed as a generic feature in many reaction-diffusion systems such as the Belousov-Zhabotinsky reaction \cite{winfree72,keener86}, 
the Fithugh-Nagumo equations  \cite{winfree91},
as well as in the amoeba system {\it Dictyostelium} \cite{palsson97}. Here a rotating spiral results from specific initial conditions, the cyclic nature of the rock-paper-scissors game, and the flux in Eq.~\ref{E_newflux} which drives players towards gain (e.g.~scissors chases paper) and away from loss (e.g.~scissors flees from rock). We also note that realizations of this game in nature, for instance the mating strategies of a California iguanid lizard \cite{sinervo96}, may be well described by our mathematical hypothesis, as such creatures cannot change their strategy.

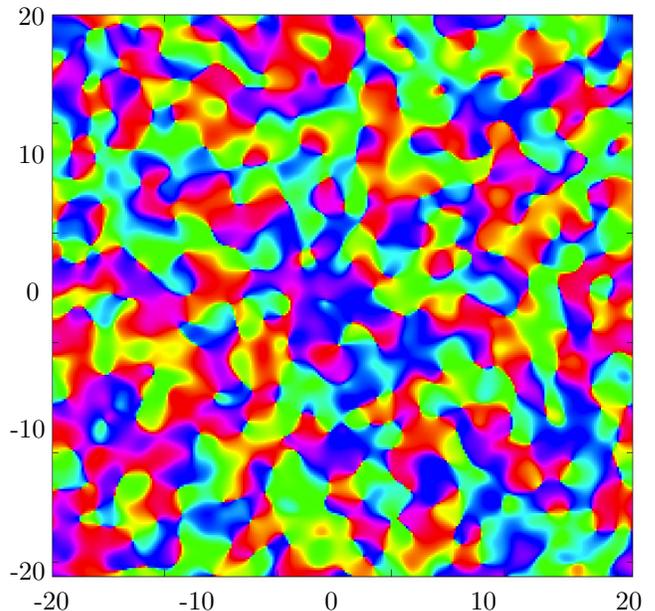
\begin{figure}[!b]
\begin{center}
\includegraphics[width=9.0cm, height=8.7cm]{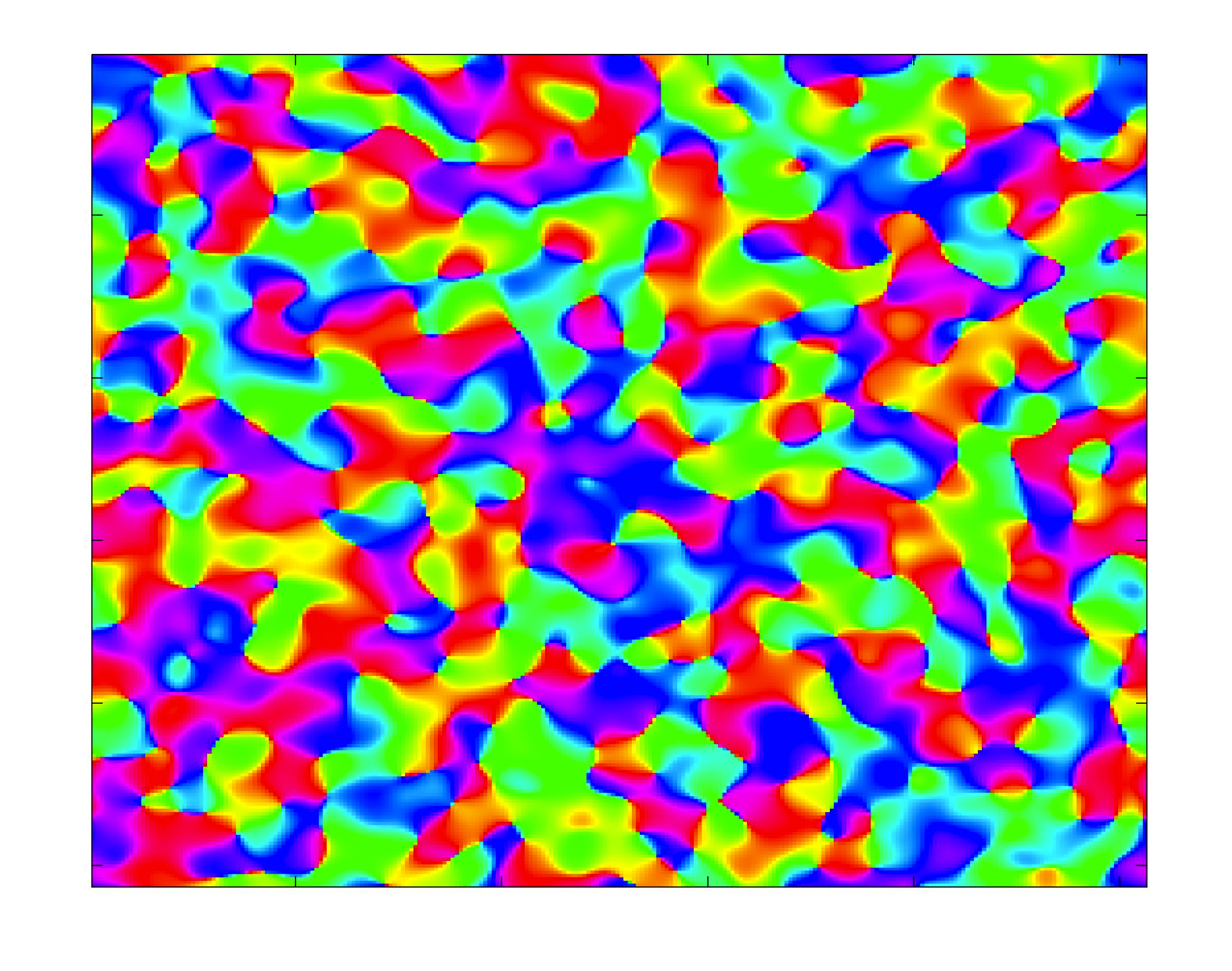} 
\put(-250,230){20}
\put(-250,178){10}
\put(-247,126){0}
\put(-253,74){-10}
\put(-253,20){-20}
\put(-244,9){-20}
\put(-189,9){-10}
\put(-134,9){0}
\put(-79,9){10}
\put(-24,9){20}
\vspace{-0.5cm}	
\caption{Disordered state in the rock-paper-scissors game, at $t = 5.0$, shown in additive RGB color.
}
\label{FIG-random}
\end{center}
\vspace{-0.35cm}	
\end{figure}

The spiral is however not a completely robust solution in our system -- it is not an attractor for all initial conditions. Fig.~\ref{FIG-random} shows a simulation at $t=5.0$ of the frozen-strategy rock-paper-scissors game, with random initial condition: the superposition of Gaussian bumps in the form of Eq.~\ref{E-bump}
centered on randomly distributed points ${\bf x}_j$. We do not observe a spontaneous organization into spiral waves, as is often seen in other systems \cite{winfree72}. Since for all finite domains we have tested, a single spiral wave eventually breaks up, Fig.~\ref{FIG-random} may be more characteristic of the asymptotic state for this spatial game. The complex pattern of predominant strategy distribution in this rock-paper-scissors game does bear a striking resemblance to the orientation map of visual selectivity in the primary visual cortex of the brain \cite{bosking97}.

In this paper we have studied the effects of a fitness gradient flux on spatial PDEs for evolutionary games with replicator dynamics, comparing the effects of this new term to diffusive coupling. We have seen numerically that the fitness gradient flux can slow or stop travelling waves in some cases, and lead to new travelling wave solutions in others. We have also introduced a PDE system with no diffusion and no strategy-changing dynamics, and studied some of the patterns that result in this frozen strategy case. The progression of initial conditions towards some final pattern characterized by a constant value of the proportionate populations $u_i/\sum u$ remains to be understood in terms of stability and basins of attraction. For frozen strategy dynamics in three strategy games (here rock-paper-scissors), the mathematical conditions for spiral stability and instability also remain to be understood.

In considering the relative motion of cells in growing embryos or tumors, a number of different aspects have been included in mathematical models of spatial motion, such as diffusion, 
chemotaxis, 
and differential adhesion \cite{glazier93,liuPNAS2011}, while
game theory has been applied to some competitive interactions \cite{grieg2003}. In this context, the equations proposed here provide a way to combine motion and competition among players (cells).
Such a directed transport may be more appropriate to spatial game theory models of biological or social systems than a random (diffusive) flux.

We would like to thank R. H. Austin for early inspiration leading to this work, and J. P. Keener, D. H. Kelley, T. Reluga, and M. J. Shelley for helpful comments and discussion.

\vspace{-0.28in}




\begin{thebibliography}{}
\bibitem{RGB}
For each strategy and at each time $t$, the value $u(x,t)$ is mapped as follows
$$
\text{Color}_u=  \frac{u-\min\{u,v,w\}}{\max\{u,v,w\}-\min\{u,v,w\}+\epsilon},
$$
and similarly for $v$ and $w$. These numbers then determine the RGB value for that
point.  The resulting image indicates which strategy is predominant,
and the relative proportion of the other two. The parameter $\epsilon$ allows the normalization to be well defined when $u=v=w$, and also acts to smooth transitions between regions with different predominant strategies. For the figures in this paper we use $\epsilon = 10^{-5}$.
\end{thebibliography}
\end{document}